\documentclass[aps,preprint]{revtex4}%
\usepackage{amsfonts}
\usepackage{amsmath}
\usepackage{amssymb}
\usepackage{graphicx}%
\providecommand{\U}[1]{\protect\rule{.1in}{.1in}}
\begin{document}
\preprint{ }
\title[Short title for running header]{Non-Riemmanian geometry, force-free magnetospheres and the generalized
Grad-Shafranov equation }
\author{Diego Julio Cirilo-Lombardo}
\affiliation{CONICET- Universidad de Buenos Aires. Instituto de Fisica del Plasma (INFIP).
Buenos Aires, Argentina, Universidad de Buenos Aires. Facultad de Ciencias
Exactas y Naturales. Departamento de Fisica. Buenos Aires, Argentina}
\affiliation{Bogoliubov Laboratory of Theoretical Physics, Joint Institute for Nuclear
Research, 141980 Dubna, Russian Federation}

\begin{abstract}
The magnetosphere structure of a magnetar is considered in the context of a
theory of gravity with dynamical torsion field beyond the standard General
Relativity (GR). To this end, the axially symmetric version of the
Grad-Shafranov equation (GSE) is obtained in this theoretical framework. The
resulting GSE\ solution in the case of the magnetosphere corresponds to a
stream function containing also a pseudoscalar part. This function solution
under axisymmetry presents a complex character that (as in the quantum field
theoretical case) could be associated with an axidilaton field.
Magnetar-pulsar mechanism is suggested and the conjecture about the origin of
the excess energy due the GSE describing the magnetosphere dynamics is
claimed. We also show that two main parameters of the electrodynamic processes
(as described in GR framework by\ Goldreich and Julian (GJ)\ in 1969
\cite{gj}) are modified but the electron-positron pair rate $\overset{\cdot
}{N}$ remains invariant. The possible application of our generalized equation
(defined in a non-Riemannian geometry) to astrophysical scenarios involving
emission of energy by gravitational waves, as described in the context of GR
in \cite{mf}, is briefly discussed.

\end{abstract}
\volumeyear{year}
\volumenumber{number}
\issuenumber{number}
\eid{identifier}
\date[Date text]{date}
\received[Received text]{date}

\revised[Revised text]{date}

\accepted[Accepted text]{date}

\published[Published text]{date}

\startpage{1}
\endpage{102}
\maketitle

\section{Introduction to the problem:}

For a long time, attempts have been made to give concrete answers to various
astrophysical and cosmological mechanisms. In particular the origin, both of
primordial fields of different types, as well as of the stellar and
cosmological dynamics. Given that both general relativity (GR) and the
standard model of elementary particles (SM) do not finish giving full
explanations to these questions, the idea of a reformulation of a unified
theory beyond RG and MS seems very attractive. In previous references, the
authors have introduced a unified model based on a non-Riemannian geometry
containing a dynamic antisymmetric torsion that admits the same results of GR
and SM already proven, but also satisfactorily solves problems that GR and SM
present difficulties or inconsistencies. Some of those problems that were
satisfactorily treated in the context of this new formulation were the
determination of the mass of the axion \cite{diego4}, violation of CP of the
neutrino \cite{diego5}, primordial magnetogenesis \cite{diego2}, etc. In this
work we calculate the equations controlling the stellar magnetospheres of
compact objects in particular, in this new context. To this end, force free
conditions are adopted by deriving the equilibrium conditions depending on a
flow function with a pseudoscalar part coming from the torsion.

Actually, a typical example is the axisymmetric force-free magnetosphere in
the exterior of a neutron star. Two possibilities are proposed for the energy
storage prior to magnetar outbursts to explain the relevant phenomena: storage
in the magnetar crust or in the magnetosphere. The latter model is discussed
in terms of similarity with solar flares (\cite{lyu}; \cite{bel}). In the
solar flare model e.g., \cite{aly}, the energy is quasi-statically stored by
thermal motion at the surface, and is suddenly released as large-scale
eruptive coronal mass ejections. The energy is dissipated via a magnetic
reconnection associated with the field reconfiguration. Analogous energy
buildup and release processes may be relevant to the magnetar giant flares,
although the energy scale differs by many orders. In sum, one must entirely
rethink the physics of neutrino cooling, photon emission, and particle
emission from a neutron star, when its magnetic field (instead of its
rotation) is the main source of free energy.This possibility is completely
feasible in the context of the model previously presented in \cite{diego3}
that is based on a geometric (Lagrangian) action that can be considered the
non-Riemannian generalization of the Born-Infeld model (see details in
\cite{diego1}\cite{diego2}\cite{diego3})%
\begin{align}
\mathcal{L}_{gs}  &  =\sqrt{\det\left[  \lambda g_{\alpha\beta}\left(
1+\frac{R_{s}}{4\lambda}\right)  +\lambda F_{\alpha\beta}\left(  1+\frac
{R_{A}}{\lambda}\right)  \right]  }\\
R_{s}  &  \equiv g^{\alpha\beta}R_{\left(  \alpha\beta\right)  };\text{
\ \ \ \ \ \ \ \ \ }R_{A}\equiv f^{\alpha\beta}R_{\left[  \alpha\beta\right]
}\text{ \ \ \ \ }%
\end{align}
$\left(  with\text{ }f^{\alpha\beta}\equiv\frac{\partial\ln\left(  \det
F_{\mu\nu}\right)  }{\partial F_{\alpha\beta}},\det F_{\mu\nu}=2F_{\mu\nu
}\widetilde{F}^{\mu\nu}\right)  $

In this model, the torsion $T_{\beta\gamma}^{\alpha}$ has a \textit{dynamic
character} (contrary to other models in the literature) and is \textit{totally
antisymmetric}, which allows it to be related to its dual vector $h_{\mu}$.
The other important feature that the energy-momentum tensor and fundamental
constants (really functions of the spacetime) are geometrically induced and
not imposed "by hand".

Field equations linking the dual vector with the electromagnetic field via the
following expression%
\begin{equation}
\nabla_{\alpha}T^{\alpha\beta\gamma}=-\lambda F^{\beta\gamma}\rightarrow
\nabla_{\left[  \beta\right.  }h_{\left.  \gamma\right]  }=-\lambda^{\ast
}F_{\beta\gamma}%
\end{equation}
which indicates that the magnetic field (in the case of interest here) is
related in this theoretical context to the dynamics of the torsion vector
$h_{\mu}$. At the same time we demonstrate, generalizing the Helmholtz theorem
in 4 dimensions, that the torsion vector admits an unique geometric
decomposition of the form
\begin{equation}
h_{\alpha}=\nabla_{\alpha}\Omega+\varepsilon_{\alpha}^{\beta\gamma\delta
}\nabla_{\beta}A_{\gamma\delta}+\gamma_{1}\overset{axial\text{ }%
vector}{\overbrace{\varepsilon_{\alpha}^{\beta\gamma\delta}M_{\beta
\gamma\delta}}}+\gamma_{2}\overset{polar\text{ }vector}{\overbrace{P_{\alpha}%
}}%
\end{equation}
where $\Omega,$ $A_{\gamma\delta},M_{\beta\gamma\delta},P_{\alpha}$fields can
be associated to particles (matter) and physical observables (e.g. vorticity,
helicity etc). In that same reference, we find via Killing-Yano symmetries,
fields and possible physical observables associated to $A_{\gamma\delta}$ and
$\Omega$ , in equation (4). In the 3 + 1 decomposition of the spacetime,
expression (4) with geometrically admissible fields (Killing-Yano symmetries)
takes the form
\begin{equation}
h_{0}=\nabla_{0}\Omega+\frac{4\pi}{3}\left[  h_{M}+q_{s}n_{s}\overline{u}%
_{s}\cdot\overline{B}\right]  +\gamma_{1}h_{V}+\gamma_{2}P_{0}%
\end{equation}%
\begin{align}
h_{i}  &  =\nabla_{i}\Omega+\frac{4\pi}{3}\left[  -\left(  \left(
\overline{A}+q_{s}n_{s}\overline{u}_{s}\right)  \times\overline{E}\right)
_{i}+\left(  \Phi+q_{s}n_{s}u_{0s}\right)  \overline{B}_{i}\right]  +\\
&  +\gamma_{1}\left[  u_{0}\left(  \overline{\nabla}\times\overline{u}\right)
+\left(  \overline{u}\times\overline{\nabla}u_{0}\right)  +\left(
\overline{u}\times\overset{\cdot}{\overline{u}}\right)  \right]  _{i}%
+\gamma_{2}P_{i}\nonumber
\end{align}
Notice that in $h_{0}$ we can recognize the magnetic and vortical helicities
where $A_{\mu}$ is the vector potential and $q_{s}$ is the particle charge,
$n_{s}$is the number density (in the rest frame) and the four-velocity of
species $s$ is $u_{s}^{\gamma}.$Consequently the simplest mechanism to
generate the necessary amount of energy of magnetospheres (even without star
rotation) can be described as follows:

1) The axion and other pseudoscalars and pseudovector particles (contained in
$h_{\alpha})$ plus all helicities increase the original magnetic field $B$
e.g.: due the induction (dynamo) linearized expression from Section II, as
\begin{equation}
\nabla\times\left(  \alpha B\right)  =h\times E-h_{0}B-\left(  E\cdot
\nabla\omega\right)  \overline{m}%
\end{equation}

2) The $B$ increased, increases the magnetic helicity $H_{M}$ defined as
($g_{3}$ determinant of the absolute space, see Section IV)%
\[
H_{M}=\int A\cdot B\sqrt{g_{3}}d^{3}x
\]

3) The $H_{M}$ in turn increases $B$ even more via expressions (3) and (7)
through the torsion vector $h_{\alpha}.$

4) Consequently, the total energy in the magnetosphere will be increased to a
certain limit.(see Section 4)%
\[
E_{M}=\int\alpha B^{2}\sqrt{g_{3}}d^{3}x
\]

5) After some limit to be determined, the excess energy in the magnetosphere
is ejected and the process is repeated.

With the above motivation, we will work out the problem of the force free
magnetosphere computing explicitly the Grad-Shafranov equation in the case of
a axisymmetric configuration (without rotation, in principle) considering the
dual of the torsion tensor $h_{\mu}$ from the gravitational theory based in
affine geometry given in \cite{diego3} . To this end the force of Lorentz in
the context of the unified model will be calculated, the 3 + 1 formalism
introduced and the geometrically induced alpha term (with introduction of the
physical currents, which intervenes in the equation of the induction producing
the dynamo effect) determined. Finally, we will present a concise discussion
on the problem of the physics of magnetospheres based on the expressions
obtained and the current knowledge regarding the intervention of high energy
processes in these scenarios.

\section{The model, generalized Lorentz force and $\alpha-$term:}

as we see before in \cite{diego1}\cite{diego2}\cite{diego3}, the geometrically
induced Lorentz force that we have been obtained from the model in the linear
limit was%
\begin{equation}
\left(  h\cdot B+\rho_{e}\right)  E+J\times B=\left(  E\cdot B\right)  h
\end{equation}
consequently, in the case of force free condition with non-vanishing torsion
field implies: $\left(  E\cdot B\right)  =0.$ General assumptions for 3+1
splitting in axisymmetrical spacetimes can be introduced in standard form
(e.g.: $j$, $E$ and $B$ can be treated as 3-vectors in spacelike
hypersurfaces). In terms of these 3-vectors the nonlinear eqs. of the original
model can be linearized and consequently written in a Maxwellian fashion as
\begin{align}
\nabla\cdot\mathbb{E} &  =-h\cdot\mathbb{B}+4\pi\rho_{e}\label{1}\\
\nabla\cdot B &  =0\label{2}\\
\nabla\times\left(  \alpha E\right)   &  =\left(  B\cdot\nabla\omega\right)
\overline{m}\label{3}\\
\nabla\times\left(  \alpha\mathbb{B}\right)   &  =h\times\mathbb{E}%
-h_{0}\mathbb{B}-\left(  \mathbb{E}\cdot\nabla\omega\right)  \overline
{m}\label{4}%
\end{align}
The derivatives in these equations are covariant derivatives with respect to
the metric of the absolute space $\gamma_{ij}$ being $\alpha,\beta:$ lapse and
shift functions respectively and $\mathbb{E=}\frac{\mathbb{\partial
}\mathcal{L}_{gs}}{\partial E}$ and $\mathbb{B=}\frac{\mathbb{\partial
}\mathcal{L}_{gs}}{\partial B}.$Because it is and unified model that we need
to replace $\overline{h}\times\overline{E}$ in order to introduce the physical
currents as follows. From the above equations in exact form, the geometrical
current induced by the non-Riemannian framework is
\begin{equation}
J\equiv\mathbb{+}h\times\mathbb{E}-h_{0}\mathbb{B}%
\end{equation}
consequently
\begin{equation}
J\times\mathbb{B\Rightarrow}\left(  h\cdot\mathbb{B}\right)  \mathbb{E=}%
J\times\mathbb{B}+\left(  \mathbb{B\cdot E}\right)  h
\end{equation}
then $h\times\mathbb{E}$
\begin{equation}
h\times\mathbb{E}=h\times\left[  \frac{\left(  \mathbb{B\cdot E}\right)
h\mathbb{+}J\times\mathbb{B}}{\left(  h\cdot\mathbb{B}\right)  }\right]
=J-\frac{\left(  h\cdot J\right)  \mathbb{B}}{\left(  h\cdot\mathbb{B}\right)
}%
\end{equation}
consequently, the relation with the physical scenario can be implemented as
follows:%
\begin{equation}
J-\frac{\left(  h\cdot J\right)  \mathbb{B}}{\left(  h\cdot\mathbb{B}\right)
}\rightarrow\alpha_{g}\left(  j_{ph}-\frac{\left(  h\cdot j_{ph}\right)
\mathbb{B}}{\left(  h\cdot\mathbb{B}\right)  }\right)
\end{equation}
transforming the set (\ref{1},\ref{2},\ref{3},\ref{4}) at the linear level,
namely $\mathbb{E}\rightarrow E$ and $\mathbb{B}\rightarrow B$, to%
\begin{align}
\nabla\cdot E &  =-h\cdot B+4\pi\rho_{e}\label{h}\\
\nabla\cdot B &  =0\\
\nabla\times\left(  \alpha_{g}E\right)   &  =\left(  B\cdot\nabla
\omega\right)  \overline{m}\\
\nabla\times\left(  \alpha_{g}B\right)   &  =\alpha_{g}\left(  j_{ph}%
-\frac{\left(  h\cdot j_{ph}\right)  B}{\left(  h\cdot B\right)  }\right)
-h_{0}B-\left(  E\cdot\nabla\omega\right)  \overline{m}\\
&  =\alpha_{g}j_{ph}-\left[  h_{0}+\frac{\left(  h\cdot j_{ph}\right)
}{\left(  h\cdot B\right)  }\right]  B-\left(  E\cdot\nabla\omega\right)
\varpi^{2}%
\end{align}

($\alpha_{g}$ is the $g_{tt}$ metrc coefficient in 3+1)From the beggining of
radio pulsar studies, three main parameters determining the key electrodynamic
processes were defined: from the calculations above, we will demonstrate in a
simple way that the quantities defined from the density are all altered. Said
alteration comes from the dynamics of $h$, being able to accentuate or even
annul the effect that the rotation has on that density. The first was the
electric charge density that is needed to screen the longitudinal electric
field near the neutron star surface, namely $\rho_{GJ}=-\frac{\Omega\cdot
B}{2\pi c},$This quantity, introduced by Goldreich and Julian (GJ)\ in 1969
\cite{gj}was used to determines the characteristic particle number density
$n_{GJ}=\frac{\left\vert \rho_{GJ}\right\vert }{\left\vert e\right\vert }$ (of
the order of 10$^{-12}$ $cm^{3}$near the neutron star surface) . Here, as h
must be considered from the equation (\ref{h}) (we concentrate on the
linearized version to simplify the analysis), the corresponding charge density
to that of GJ is
\[
\rho_{UFT}=-\frac{\left(  \Omega+h\right)  \cdot B}{2\pi c}\equiv\rho
_{GJ}+\rho_{h}%
\]
(subindices GJ indicate here the corresponding GJ quantity) consequently the
characteristic charge density can only be determined through the knowledge of
$h$, and the corresponding characteristic number density that will be
\[
n_{UFT}=\left\vert -\frac{\left(  \Omega+h\right)  \cdot B}{2\pi c}\right\vert
/\left\vert e\right\vert
\]
Also the characteristic current density, is modified as%
\[
j_{UFT}=c\rho_{UFT}%
\]
which is much more important as indicated in \cite{lov} because in such
approaches it is the longitudinal electric current circulating in the
magnetosphere that will play the key role.

The second parameter is the particle multiplication defined currently as
$\lambda_{GJ}=n_{e}/n_{GJ}$, which shows how much the secondary particle
number density exceeds the critical number density $n_{GJ}$. Also this
parameter is affected according to our work, as%
\[
\lambda_{UFJ}=n_{e}/n_{UFT}%
\]
that is evidently greater than the same GJ\ quantity. As inside of the above
expression we have $n_{h}$, the secondary particle number density must be
greater than in the GJ\ case to exceeds the new critical number density
$n_{UFT}$. Finally, the third relevant quantity is the hydrodynamic particle
flow that now is $\overset{\cdot}{N}_{UFT}m_{e}c^{2}\Gamma$ ( $\Gamma$ here
and below denotes the hydrodynamic Lorentz factor of the outflowing plasma)
with the\ electron-positron pair injection rate
\[
\overset{\cdot}{N}_{UFT}=c\pi\lambda_{UFT}R_{0}^{2}n_{UFT}=\overset{\cdot}{N}%
\]
that, with these definitions, it is not modified.

\section{Force free magnetospheres: generalized Grad-Shafranov equation:}

\textit{ }The consistent theoretical description of gravitational
magnetohydrodynamics (MHD) equilibria is of fundamental importance for
understanding the phenomenology of accretion disks (AD) around compact objects
(black holes, neutron stars, etc.). The very existence of these equilibria is
actually suggested by observations, which not only show evidence of quiescent,
and essentially non-relativistic, AD plasmas close to compact stars, also the
dynamical interplay with high energy processes involving the magnetospheres of
compact objects, in particular pulsars, quasars and magnetars. The
electromagnetic (EM) fields involved, in particular the electric field, may
locally be extremely intense\cite{gj}, so several standard processes such as
electron positron pair creation occur, but several exotic interactions
involving neutrinos with axions and other dark matter candidates must also be
taken into account. This suggests therefore that such equilibria (if it
certainly exists) should be described in the framework of unified field theory
beyond general relativity (GR)\cite{diego1}\cite{diego2}\cite{diego3}%
\cite{diego4} and beyond the standard model (SM). Extending previous
approaches, holding for compact objects/black hole axisymmetric geometries
having into account effect of space-time curvature, the purpose of this work
is the formulation of a generalized Grad-Shafranov (GGS)\cite{grad}%
\cite{sha}\cite{solo}\cite{za} equation based in a non-Riemannian geometry
with dynamical torsion field suitable for the investigation of accretion, jets
and winds and other astrophysical effects when high energy effects (exotic or
not)are present. Now we will calculate the GSE with axisymmetry. Arguments and
procedures for calculating GSE in this model are similar in form to works well
known in the context of GR\cite{McTh}\cite{oka}\cite{lov} (we use through the
work\cite{oka} notation) to have a reasonable comparison parameter with those results.

\section{Magnetic fields and currents:\textit{ }}

From the electrodynamic equations in 3+1 formulation of curved spacetimes,
under axisymmetry, the magnetic field is spplitted as $\overline{B}%
=B_{p}+B_{t}$ where%

\begin{align}
B_{p}  &  =\frac{1}{2\pi\varpi^{2}}\left(  \nabla\psi+\overline{m}\times\nabla
h_{0}\right)  \times\overline{m}=\frac{1}{2\pi\varpi}\left(  \nabla\psi\times
e_{\widehat{\phi}}+\varpi\nabla h_{0}\right)  =\frac{\nabla\psi\times
\overline{m}}{2\pi\varpi^{2}}+\nabla\chi\label{s}\\
B_{t}  &  =-\frac{2I\left(  \psi,\chi\right)  }{\alpha\varpi c}e_{\widehat
{\phi}}\nonumber
\end{align}
with $e_{\widehat{\phi}}$ unitary toroidal vector, $h_{0}$ pseudoscalar field
that we redefine as $\chi$(zero component of the dual of the antisymmetric
torsion field) and $\overline{m}\cdot\overline{m}=g_{\phi\phi}=\varpi^{2}.$
The expression for the toroidal magnetic field coming from the Ampere law and
the currents enclosed by the surface $A$ , namely $I$ (depending on $\psi$ and
$\chi),$ are obtained similarly to the magnetic flux assuming the form:
$\nabla I=\nabla I\left(  \psi\right)  +\overline{m}\times\nabla I\left(
\chi\right)  $ $.$ Consequently, due that $\left(  E\cdot B\right)  =0$ the
eqs.(\ref{s}) brings the force free condition as%

\begin{align}
j_{p}  &  =\frac{1}{2\pi\alpha\varpi}\left(  e_{\widehat{\phi}}\times\nabla
I\right)  =-\frac{\nabla I\left(  \psi\right)  \times\overline{m}}{2\pi
\alpha\varpi^{2}}+\frac{\nabla I\left(  \chi\right)  }{2\pi}\\
&  =-\frac{1}{\alpha}\frac{dI}{d\zeta}B_{p}\nonumber
\end{align}
where we have defined the multi-vector $\zeta\equiv\psi+\overline{m}\chi$
$\left(  \overline{m}\equiv\varpi e_{\widehat{\phi}}\right)  $ (and
consequently under action of exterior derivative: $d\zeta=d\psi+\overline
{m}\times d\chi$ and $\nabla\zeta=\nabla\psi+\overline{m}\times\nabla\chi).$ Thus:%

\begin{equation}
B_{p}\cdot dA=d\zeta
\end{equation}%
\begin{equation}%
%TCIMACRO{\doint }%
%BeginExpansion
{\displaystyle\oint}
%EndExpansion
\alpha j_{p}\cdot dA=-%
%TCIMACRO{\doint }%
%BeginExpansion
{\displaystyle\oint}
%EndExpansion
\frac{dI}{d\zeta}B_{p}\cdot dA=-%
%TCIMACRO{\dint \limits_{0}^{\zeta}}%
%BeginExpansion
{\displaystyle\int\limits_{0}^{\zeta}}
%EndExpansion
\frac{dI}{d\zeta}B_{p}\cdot dA
\end{equation}%
\begin{equation}
I\left(  0\right)  =I\left(  \zeta\right)
\end{equation}%
\begin{equation}
j_{t}=-\frac{1}{8\pi}\left[  \frac{\varpi c}{\alpha}\nabla\cdot\left(
\frac{\alpha}{\varpi^{2}}\underset{\equiv\nabla\zeta}{\underbrace{\left(
\nabla\psi+\overline{m}\times\nabla\chi\right)  }}\right)  +\frac
{\varpi\left(  \Omega_{F}-\omega\right)  }{\alpha^{2}c}\left(  \nabla
\psi+\overline{m}\times\nabla\chi\right)  \cdot\nabla\omega\right]  \label{a}%
\end{equation}%
\begin{equation}
v_{F}=\frac{1}{\alpha}\left(  \Omega_{F}-\omega\right)  \varpi
\end{equation}%
\begin{equation}
E_{p}=-\frac{v_{F}}{c}\left(  e_{\widehat{\phi}}\times B_{p}\right)
=-\frac{1}{2\pi c\alpha}\left(  \Omega_{F}-\omega\right)  \nabla\zeta\text{
(force free)}%
\end{equation}%
\begin{align}
\rho_{e}  &  =\frac{1}{4\pi}\left(  \nabla\cdot E_{p}+\overline{h}\cdot
B\right) \\
&  =\frac{1}{4\pi}\nabla\cdot E_{p}\text{ \ \ }\nonumber
\end{align}
then%
\begin{equation}
\rho_{e}=-\frac{1}{8\pi^{2}}\frac{\varpi^{2}\left(  \Omega_{F}-\omega\right)
}{\alpha^{2}c}\left\{  \nabla\cdot\left(  \frac{\alpha}{\varpi^{2}}\nabla
\zeta\right)  +\frac{\alpha}{\varpi^{2}}\nabla\zeta\cdot\nabla\ln\left[
\frac{\varpi^{2}\left(  \Omega_{F}-\omega\right)  }{\alpha^{2}c}\right]
\right\}  \label{b}%
\end{equation}
as usual we can eliminate the first factor: $\nabla\cdot\left(  \frac{\alpha
}{\varpi^{2}}\left(  \nabla\Psi+\overline{m}\times\nabla\chi\right)  \right)
$ between expressions (\ref{a}) and (\ref{b})$,$ consequently%
\begin{equation}
\rho_{e}-\frac{\varpi\left(  \Omega_{F}-\omega\right)  }{\alpha c}\frac{j_{t}%
}{c}=-\frac{1}{8\alpha\pi^{2}}\left\{  \left(  \frac{\varpi\left(  \Omega
_{F}-\omega\right)  }{\alpha c}\right)  ^{2}\nabla\zeta\cdot\nabla\omega
-\frac{c\alpha^{2}}{\varpi^{2}}\nabla\zeta\cdot\nabla\left(  \frac{\varpi
^{2}\left(  \Omega_{F}-\omega\right)  }{\alpha^{2}c}\right)  \right\}
\end{equation}
In this case with dynamical torsion field, the transfield component of the
momentum equation for the force free case, namely%
\begin{equation}
\left(  \overline{h}\cdot B+\rho_{e}\right)  E+J\times B=0
\end{equation}
becomes to%
\begin{equation}
\frac{j_{t}}{c}-\frac{\varpi\left(  \Omega_{F}-\omega\right)  }{\alpha c}%
\rho_{e}=\frac{1}{2\alpha^{2}\varpi c^{2}}\frac{IdI}{d\zeta}%
\end{equation}
Solving for $\rho_{e}$ and $j_{t}$%
\begin{equation}
8\pi^{2}\rho_{e}=\frac{\frac{\varpi\left(  \Omega_{F}-\omega\right)  }{\alpha
c}}{1-\left(  \frac{\varpi\left(  \Omega_{F}-\omega\right)  }{\alpha
c}\right)  ^{2}}\left[  \frac{8\pi^{2}}{2\alpha^{2}\varpi c}\frac{IdI}{d\zeta
}+\frac{\varpi\left(  \Omega_{F}-\omega\right)  }{\alpha^{2}c}\nabla\zeta
\cdot\nabla\omega-\frac{c}{\varpi}\nabla\zeta\cdot\nabla\ln\left(
\frac{\varpi^{2}\left(  \Omega_{F}-\omega\right)  }{\alpha^{2}c}\right)
\right]  \label{r}%
\end{equation}%
\begin{align}
8\pi^{2}j_{t}  &  =\frac{1}{1-\left(  \frac{\varpi\left(  \Omega_{F}%
-\omega\right)  }{\alpha c}\right)  ^{2}}\left[  \frac{8\pi^{2}}{2\alpha
^{2}\varpi c}\frac{IdI}{d\zeta}+\frac{\varpi\left(  \Omega_{F}-\omega\right)
}{\alpha^{2}c}\nabla\zeta\cdot\nabla\omega-\frac{c}{\varpi}\nabla\zeta
\cdot\nabla\ln\left(  \frac{\varpi^{2}\left(  \Omega_{F}-\omega\right)
}{\alpha^{2}c}\right)  \right] \label{j}\\
&  (\text{ it is due Ampere eq. (fourth Maxwell eq. above)projected
toroidally)}\nonumber
\end{align}
Consequently, from(\ref{r}) and (\ref{j}) we obtain%
\begin{equation}
\nabla\cdot\left(  \frac{\alpha}{\varpi^{2}}\nabla\zeta\right)  =\nabla
\cdot\left(  \frac{\left(  \Omega_{F}-\omega\right)  ^{2}}{\alpha c^{2}}%
\nabla\zeta\right)  -\frac{\left(  \Omega_{F}-\omega\right)  }{\alpha c^{2}%
}\frac{d\Omega_{F}}{d\zeta}\left\vert \nabla\zeta\right\vert ^{2}-\frac
{8\pi^{2}}{2\alpha\left(  \varpi c\right)  ^{2}}\frac{IdI}{d\zeta} \label{k}%
\end{equation}

Eqs.(\ref{r}),(\ref{j})and (\ref{k}) are the Grad Shafranov ones, notice that
which can be seen as a 2 dimensional Poisson type equation (axisymmetry) of a
complex variable $\zeta\equiv\psi+\overline{m}\chi\rightarrow\psi+i\chi$ with
a source term $\varpropto$ $\frac{IdI}{d\zeta}$

\section{Discussion}

In order to help te reader and on the possible physical scenario in which we
could study our model with respect to the energy limits and the possible
emission mechanisms, a good analysis using the Post-Newtonian method in the
context of GR standard was carried out in reference \cite{mf}.

In \cite{mf} reference they have calculated the electromagnetic corrections to
the gravitational waves emitted by a coalescing binary system as a
contribution to the total enegy-momentum tensor (EMT) of a dipolar
electromagnetic field. Consequently the goal in that case was the
determination of the correction to the emission of standard gravitational
energy by a gravitomagnetic term that becomes null when the magnetic field
becomes zero.

In our case (as it is easy to see in references\cite{diego2} \cite{diego3})
the source of the gravitational field and the electromagnetic field are given,
in a unified way, by the geometrically induced EMT then will be very
interesting in a future work to develop in the context of our model, the same
procedure as in \cite{mf} to study the same physical scenario.

\section{Concluding remarks and outlook:}

\textit{ }As we saw in previous works, the physical currents are linked to the
torsion vector h by means of its only decomposition in fields of matter
(particles) and observables. Being a pseudoscalar field playing the role of
axion and magnetic, vortex and mixed helicities respectively. Also $P_{0}$ an
arbitrary polar vector with $\gamma_{2}$ pseudoscalar quantity that we will
put equal to zero, in principle. Notice that geometrically the vector torsion
field can be uniquely decomposed as
\begin{equation}
h_{0}=\nabla_{0}a+\varepsilon_{\alpha}^{\gamma\delta\rho}\frac{4\pi}{3}\left[
h_{M}+q_{s}n_{s}\overline{u}_{s}\cdot\overline{B}\right]  +\gamma_{1}%
h_{V}+\gamma_{2}P_{0}%
\end{equation}
being $a$ pseudoscalar field playing the role of axion and $h_{M}$ magnetic,
$h_{V}$ vortex($\gamma_{1}$ scalar) and $\overline{u}_{s}\cdot\overline{B}$
mixed helicities respectively. Also $P_{0}$ an arbitrary polar vector with
$\gamma_{2}$ pseudoscalar quantity that we will put equal to zero, in
principle. Consequently, the complex flow function contains the dynamics of
the axion (candidate of dark matter) and the helicities corresponding to the
term alpha in the equation of induction that generates the astrophysical
dynamo effect e.g.:%
\begin{equation}
\nabla\zeta=\nabla\psi+\overline{m}\times\nabla\underset{h_{0}}{\underbrace
{\left(  \nabla_{0}a+\varepsilon_{\alpha}^{\gamma\delta\rho}\frac{4\pi}%
{3}\left[  h_{M}+q_{s}n_{s}\overline{u}_{s}\cdot\overline{B}\right]
+\gamma_{1}h_{V}+\gamma_{2}P_{0}\right)  }}%
\end{equation}

($a$ is the axion field). It is interesting to note that, in contrast to this
work that is of first principles, the helicities in the alpha term that causes
the anomalous current proportional to the magnetic field $B$ were suggested in
recent works of astrophysics\cite{hel} (pulsars, magnetars, gravastars) and
placed "by hand"

Points to be considered in future work will be the different types of
accretion with dark matter and effects in jets and mechanisms of accretion in
pulsars and effects of emission of gravitational waves in compact objects and
black holes. Also, the exotic interactions and charge separation due to the
pseudovectorial character of the torsion field $h_{\mu}$.

\section{Acknowledgements:}

This work is devoted to a friend and excellent colleage Nikolay Kochelev. We
are very grateful to Professor Max Camenzind for bring us some detaills of the
computations and features of the force free approach in magnetospheres of
compact objects. Also we are grateful to CONICET-Argentina for finnantial
support and BLTP-JINR\ for hospitality and institutional support

\end{document}